\documentclass{JHEP3} 
\usepackage{amsmath}

\title{Calculating the gauge-invariant two-point correlation function of gluonic field strengths}

\author{Dmitri Antonov
        \thanks{Permanent address: 
        ITEP, B. Cheremushkinskaya 25, RU-117 218 Moscow, Russia.}\\ 
        Institute of Physics, Humboldt University of Berlin,\\
        Newtonstr. 15, 12489 Berlin, Germany \\       
        E-mail: \email{antonov@physik.hu-berlin.de}}

\abstract{The two-point gauge-invariant correlation function of gluonic field strengths,
which is the main input in the stochastic vacuum model, is derived by using its relation to the 
Green functions of one- and two-gluon gluelumps. These Green functions are found by evaluating the 
path integrals of one and two gluons confined in the field of an infinitely heavy adjoint source. This 
allows one to obtain also the  
terms in the exponential asymptotics of the two-point correlation function, which are produced by fluctuations
of the gluons around their mean separation from the source.
The pre-exponents are derived as well, demonstrating some novel dependences on the distance.}


\preprint{HU-EP-05/70} 

\dedicated{}

\begin{document}

\section{Introduction}
Despite a remarkable impact of the stochastic vacuum model~\cite{cdm, svm} on QCD phenomenology (see e.g.~\cite{book} for a recent 
review), a field-theoretic establishment of this model is still pending. The main problem is an analytic calculation
of the nonperturbative content of the two-point gauge-invariant correlation function of gluonic field strengths, which can be 
parametrized as~\cite{svm}
$$\frac{g^2}{N_c}{\rm tr}{\,}\left<F_{\mu\nu}(x)\Phi_{xx'}F_{\lambda\rho}(x')\Phi_{x'x}\right>_{\rm YM}=
(\delta_{\mu\lambda}\delta_{\nu\rho}-\delta_{\mu\rho}\delta_{\nu\lambda})D(x-x')+$$
\begin{equation}
\label{cf}
+\frac12\left[\partial_\mu^x\left((x-x')_\lambda\delta_{\nu\rho}-(x-x')_\rho\delta_{\nu\lambda}\right)+
\partial_\nu^x\left((x-x')_\rho\delta_{\mu\lambda}-(x-x')_\lambda\delta_{\mu\rho}\right)\right]D_1(x-x').
\end{equation}
Here, $\Phi_{xx'}\equiv {\cal P}{\,}\exp\left(ig\int_{x'}^{x}dz_\mu A_\mu^a(z)T_{\rm fund}^a\right)$ is a phase 
factor along the straight line, which goes through $x'$ and $x$; $T_{\rm fund}^a$'s
stand for the generators of the group SU($N_c$) in the fundamental representation; the average $\left<\ldots\right>_{\rm YM}$
is taken with respect to the Euclidean Yang-Mills action, $\frac14\int d^4x(F_{\mu\nu}^a)^2$; 
$F_{\mu\nu}^a=\partial_\mu A_\nu^a-\partial_\nu A_\mu^a+gf^{abc}A_\mu^b A_\nu^c$, $a=1, \ldots, N_c^2-1$.

Numerous lattice measurements of the correlation function~(\ref{cf}) were performing over the years (see~\cite{cdm, d} for the calculations 
with the cooling method and ref.~\cite{nga}, where this correlation function was reconstructed from the lattice data on the 
static quark-antiquark potential). As for 
the papers attempting to analytically evaluate this quantity, some of those were dealing with the 
Abelian theories, where confinement was produced by monopoles~\cite{early, later}. At the same time, some progress was achieving in the
non-Abelian case as well~\cite{dm, na}. In particular, a relation between the correlation length of the vacuum (that is the 
length at which the functions $D$ and $D_1$ experience an exponential fall-off) 
and the gluonic condensate was derived in~\cite{dm}. Further analysis of the correlation lengths of $D$ and $D_1$
was done in~\cite{na}.~\footnote{It was supporting the conjecture, tested on the lattice in~\cite{nga}, that these lengths might be different.}
Recently, a complete derivation of these functions was presented in~\cite{S}. Using initially
the idea~\cite{dm} on a relation between the correlation function~(\ref{cf}) and the Green functions 
of one and two (valence) gluons confined in the field of an infinitely heavy adjoint source (the so-called one- and two-gluon gluelumps),
the paper~\cite{S} further accommodated the Hamiltonian formalism for the analysis 
of these Green functions. This eventually yielded certain nonperturbative expressions for the functions $D$ and $D_1$.

In the present paper, the Green functions (as well as the functions $D$ and $D_1$) will be calculated upon an evaluation 
of the corresponding integrals over trajectories of the gluons in the gluelumps. It turns out that, when the minimal-area-law
Ansatz is applied to the Wilson loops of these (confined) gluons, further calculation of the path integrals 
becomes possible by making use of the Feynman variational method. By means of this method, we manage to find
some novel terms in both the exponents and pre-exponents of the functions $D$ and $D_1$, which stem from 
the vibrations of the gluons around their mean separations from the heavy source. 

The paper is organized as follows. In the next section, we demonstrate our method at the example of a one-gluon gluelump and 
calculate in this way the function $D_1$. In section~3, we proceed with the two-gluon gluelump and the function $D$. 
In section~4, a brief summary of the obtained results is presented.

\section{Green function of a one-gluon gluelump and the function $D_1$}
In the approximation when the gluon spin term is disregarded~\cite{S}, 
the one-gluon gluelump Green function has the form $\delta_{\mu\nu}G(T,0)$, where
$$G(T,0)=\int\limits_{0}^{\infty}ds\int\limits_{x_4(0)=0,{\,}x_4(s)=T}^{}{\cal D}x_\mu
\exp\left(-\int_{0}^{s}d\tau\frac{\dot x_\mu^2}{4}\right)\left<W(C)\right>_{\rm YM}.$$
Here, in the Wilson loop $\left<W(C)\right>_{\rm YM}={\rm tr}\left<{\cal P}{\,}\exp\left(ig\oint_{C}^{}dx_\mu A_\mu^aT_{\rm adj}^a\right)\right>_{\rm YM}$,
$T_{\rm adj}^a$'s stand for the SU($N_c$)-generators in the adjoint representation, $(T_{\rm adj}^a)^{bc}=-if^{abc}$; the closed contour $C$ 
is formed by the piece of the 4-th axis from the point $({\bf 0}, T)$ to the origin and the trajectory $x_\mu(\tau)$ of a gluon, which
moves back from the origin to that point. Adopting for $\left<W(C)\right>_{\rm YM}$ the minimal-area-law Ansatz, we have
$$
G(T,0)=N_c(N_c^2-1)\int\limits_{0}^{\infty}ds\int\limits_{x_4(0)=0,{\,}x_4(s)=T}^{}{\cal D}x_4\int{\cal D}{\bf x}\times$$
$$\times\exp\left[-\int_{0}^{s}d\tau\left(\frac{\dot{\bf x}^2}{4}+\frac{\dot x_4^2}{4}+\sigma_{\rm adj}|{\bf x}|\dot x_4
\right)\right].$$
It will further be assumed that, in the directions transverse to $x_4$, gluon, along most of its path, only slightly fluctuates  
around some mean distance, $L$, from the 4-th axis.~\footnote{This distance will be found numerically at the end of the calculation.} 
This means the following parametrization ${\bf x}(\tau)=
{\bf L}+{\bf l}(\tau)$, where ${\bf l}(0)={\bf l}(s)=0$, ${\bf L}=L{\bf n}$. Then, up to the term 
linear in ${\bf l}$, $|{\bf x}|\simeq L+{\bf n}{\bf l}$. In this approximation, we have
$$
G(T,0)\simeq N_c(N_c^2-1){\rm e}^{-\sigma_{\rm adj}LT}
\int\limits_{0}^{\infty}ds\int\limits_{x_4(0)=0,{\,}x_4(s)=T}^{}{\cal D}x_4\int{\cal D}{\bf l}{\cal D}\mu\times$$
$$\times\exp\left[-\int_{0}^{s}d\tau\left(\frac{\dot{\bf l}^2}{4}+\frac{\dot x_4^2}{4}+\frac{\mu}{2}+
\frac{\sigma_{\rm adj}^2\dot x_4^2({\bf n}{\bf l})^2}{2\mu}\right)\right],$$
where we have introduced an integration over the ``einbein'' field $\mu$ in order to make the potential term 
quadratic in ${\bf l}$. Next,
the unit vector ${\bf n}$ may, without loss of generality, be chosen along the 1-st axis, ${\bf n}=(1,0,0)$.
The integrations over $l_2(\tau)$ and $l_3(\tau)$ then become those of a free particle. As for the integration over $l_1(\tau)$, 
since ${\bf l}$, by definition, 
only weakly varies along the trajectory, one may replace $l_1^2(\tau)$ by its value averaged along the trajectory, 
$l_1^2\to\frac{1}{s}\int_{0}^{s}d\tau l_1^2$. Altogether we have
$$
G(T,0)\simeq N_c(N_c^2-1){\rm e}^{-\sigma_{\rm adj}LT}
\int\limits_{0}^{\infty}\frac{ds}{4\pi s}
\int\limits_{x_4(0)=0,{\,}x_4(s)=T}^{}{\cal D}x_4\int{\cal D}l_1{\cal D}\mu\times$$
\begin{equation}
\label{G}
\times\exp\left[-\int_{0}^{s}d\tau\left(\frac{\dot l_1^2}{4}+\frac{\dot x_4^2}{4}+\frac{\mu}{2}+
U(l_1)\right)\right],~~ U(l_1)=\left(\frac{\sigma_{\rm adj}^2}{2s}\int_{0}^{s}d\tau\frac{\dot x_4^2}{\mu}\right)
l_1^2.
\end{equation} 
Although the partition function of a harmonic oscillator, resulting now from the integration over $l_1(\tau)$, is known exactly, 
its substitution would be of little use for a practical calculation. (Merely since the frequency of this oscillator 
is a functional of $\dot x_4/\mu$, and the subsequent integrals over $x_4$ and $\mu$ can hardly be handled.)
Rather, it is much more useful to apply
Feynman variational method~\cite{2}, that is fully legitimate, since $l_1$ only slightly fluctuates 
around the origin. This method yields the leading quantum correction to the classical expression 
\begin{equation}
\label{clasint}
\frac{1}{\sqrt{4\pi s}}{\rm e}^{-sU(0)}
\end{equation} 
for the integral $\int\limits_{x(0)=x(s)=0}^{}{\cal D}x(\tau)\exp\left[-\int_{0}^{s}d\tau
\left(\frac{\dot x^2}{4}+U(x)\right)\right]$. The correction originates just from the fluctuations
of the trajectory around the origin and leads to the following expression for the above-mentioned integral (see e.g.~\cite{sb} for the details of derivation): 
$\frac{1}{\sqrt{4\pi s}}\exp\left[-\sqrt{\frac{3s}{\pi}}\int_{-\infty}^{+\infty} dxU(x){\rm e}^{-\frac{3x^2}{s}}\right]$.
The classical approximation~(\ref{clasint}) is recoverable from this formula in the limit $s\to 0$.
 
Substituting for $U$ the expression from eq.~(\ref{G}), we then get for  
the path integral over $l_1(\tau)$: $\frac{1}{\sqrt{4\pi s}}\exp\left(-\frac{\sigma_{\rm adj}^2s}{12}\int_{0}^{s}
d\tau\frac{\dot x_4^2}{\mu}\right)$. This yields
$$G(T,0)= N_c(N_c^2-1){\rm e}^{-\sigma_{\rm adj}LT}
\int\limits_{0}^{\infty}\frac{ds}{(4\pi s)^{3/2}}
\int\limits_{x_4(0)=0,{\,}x_4(s)=T}^{}{\cal D}x_4{\cal D}\mu\times$$
\begin{equation}
\label{intl}
\times\exp\left[-\int_{0}^{s}d\tau\left(\frac{\dot x_4^2}{4}+\frac{\mu}{2}+
\frac{\sigma_{\rm adj}^2s}{12}\frac{\dot x_4^2}{\mu}\right)\right].
\end{equation}
Further integration over $\mu$ replaces the sum $\frac{\mu}{2}+\frac{\sigma_{\rm adj}^2s}{12}\frac{\dot x_4^2}{\mu}$
by a single term $\sigma_{\rm adj}\sqrt{s/6}|\dot x_4|$. This is a world-line Lagrangian of a free scalar particle, which moves along the 4-th axis.  
(The mass of this particle $m\equiv\sigma_{\rm adj}\sqrt{s/6}$ is, however, a function of $s$.)
Such a Lagrangian is known to be equivalent to the one usually used for the world-line calculations, $\frac{\dot x_4^2}{4}+m^2$,
so that the total world-line action reads $\int_{0}^{s}d\tau\frac{\dot x_4^2}{2}+m^2s$. To restore the standard coefficient 
1/4 at the kinetic term, one can perform the rescaling $\tau=2\tau_{\rm new}$, $s=2s_{\rm new}$. The integral over $x_4$
then becomes standard, and we obtain 
$$
G(T,0)=N_c(N_c^2-1)\frac{{\rm e}^{-\sigma_{\rm adj}LT}}{16\sqrt{2}\pi^2}\int\limits_{0}^{\infty}\frac{ds}{s^2}
\exp\left(-\frac{T^2}{4s}-\frac{2\sigma_{\rm adj}^2}{3}s^2\right).
$$
To calculate the integral over $s$, let us change the integration variable to $\xi=(\sigma s)^{-1}$, that yields for this integral 
$\sigma\int_{0}^{\infty}d\xi\exp\left(-\frac{\sigma_{\rm adj}T^2}{4}\xi-\frac{2}{3\xi^2}\right)$. The saddle-point 
approximation to the integral $\int_{0}^{\infty}d\xi {\rm e}^{-f(\xi)}$ 
works well when the width of the corresponding Gaussian distribution,
which is $\frac{1}{\sqrt{f''(\xi_0)}}$ (where $\xi_0$ is the saddle point), 
is much smaller than 1. In our case, $\frac{1}{\sqrt{f''(\xi_0)}}=
(32/9)^{1/3}(\sigma_{\rm adj}T^2)^{-2/3}$, 
therefore the saddle-point approximation is legitimate when the distances under consideration are large in the sense of the inequality 
$\sigma_{\rm adj}T^2\gg 1$. Calculation of the integral is then straightforward, and we have 
$$
G(T,0)=N_c(N_c^2-1)\frac{(\sigma_{\rm adj}/T^4)^{1/3}}{3^{2/3}\cdot 2^{7/3}\cdot \pi^{3/2}}
\exp\left[-\sigma_{\rm adj}LT-(9/32)^{1/3}(\sigma_{\rm adj}T^2)^{2/3}\right].$$
The function $D_1(x)$ is related to this expression as~\cite{S} $D_1(x)=-\frac{2g^2}{N_c^2}\frac{dG(x,0)}{dx^2}$.  
Performing the differentiation we obtain
$$
D_1(x)=\frac{C_2\alpha_s}{\sqrt{\pi}}(32/9)^{1/3}\frac{\sigma_{\rm adj}}{x^2}\left[\frac{1}{12^{1/3}}+\frac{L}{2}
\left(\frac{\sigma_{\rm adj}}{|x|}\right)^{1/3}+\frac{2}{3(\sigma_{\rm adj}x^2)^{2/3}}\right]\times$$
\begin{equation}
\label{D1}
\times\exp\left[-\sigma_{\rm adj}L|x|-(9/32)^{1/3}(\sigma_{\rm adj}x^2)^{2/3}\right]~ {\rm at}~ \sigma_{\rm adj}x^2\gg 1,
\end{equation}
where $C_2=\frac{N_c^2-1}{2N_c}$ is the quadratic Casimir operator of the fundamental representation.
Note that a gluon is confined only at large $N_c$, where $\sigma_{\rm adj}=2\sigma_{\rm fund}$ (while, 
at $N_c\sim 1$, it is screened by other gluons). In the large-$N_c$ limit, $C_2\alpha_s\to\frac{\lambda}{8\pi}$, where 
$\lambda=g^2N_c$ is the so-called 't~Hooft coupling, which stays finite in this limit.

Equation~(\ref{D1}) can further be compared to the expression for $D_1(x)$ 
found in ref.~\cite{S} by the Hamiltonian approach, which reads 
\begin{equation}
\label{D1S}
D_1(x)=\frac{2C_2\alpha_sM_0\sigma_{\rm adj}}{|x|}{\rm e}^{-M_0|x|}~ {\rm valid}~ {\rm at}~ M_0|x|\gg 1,
\end{equation} 
where $M_0\simeq 1.5{\,}{\rm GeV}$ at $\sigma_{\rm fund}=0.18{\,}{\rm GeV^2}$.
One can see that the term $(9/32)^{1/3}(\sigma_{\rm adj}x^2)^{2/3}$ in eq.~(\ref{D1}), although parametrically dominating the exponent  
in the limit $\sigma_{\rm adj}x^2\gg 1$, is numerically smaller than the leading term $\sigma_{\rm adj}L|x|$.
This is not surprising, since the term $(9/32)^{1/3}(\sigma_{\rm adj}x^2)^{2/3}$, originating 
from small fluctuations of the valence gluon around its mean separation from the heavy source, can only be a correction
to the leading result. Indeed, equating $\sigma_{\rm adj}L$ to $M_0$, we have 
$\frac{(9/32)^{1/3}(\sigma_{\rm adj}x^2)^{2/3}}{\sigma_{\rm adj}L|x|}=
\frac{\left(9\sigma_{\rm fund}^2|x|\right)^{1/3}}{2M_0}$.
Next, in the pure Yang-Mills theory which we consider here, 
the stochastic vacuum model is only applicable up to the distances where the glueball production starts. This happens
at $|x|_{\rm max}=\frac{M_{\rm glueball}}{\sigma_{\rm fund}}$. Setting for an estimate $M_{\rm glueball}=1{\,}{\rm GeV}$, 
we get $|x|_{\rm max}=5.56{\,}{\rm GeV}^{-1}$. (This yields $\sigma_{\rm adj}|x|_{\rm max}^2=11.11$, that still leaves 
lot of space for the inequality $\sigma_{\rm adj}x^2\gg 1$ to hold.) Accordingly, $\frac{\left(9\sigma_{\rm fund}^2|x|\right)^{1/3}}{2M_0}\le 0.39$,
that demonstrates the numerical smallness of the term 
$(9/32)^{1/3}(\sigma_{\rm adj}x^2)^{2/3}$ with respect to the term $\sigma_{\rm adj}L|x|$.

Let us finally consider the pre-exponent of eq.~(\ref{D1}). First, the relation $\sigma_{\rm adj}L=M_0$ yields 
$L=0.92{\,}{\rm fm}$. This value for the mean distance, by which a valence gluon  
is separated from an infinitely heavy adjoint source, well exceeds the
corresponding vacuum correlation length $M_0^{-1}\simeq0.15{\,}{\rm fm}$.
The string in a gluelump is therefore a clearly shaped object. One can further see that, at 
this value of $L$, the term $\frac{1}{12^{1/3}}$ in the brackets is always larger than the term  
$\frac{L}{2}\left(\frac{\sigma_{\rm adj}}{|x|}\right)^{1/3}$.
Indeed, to compare the latter to the former means to compare 
$\frac{\sigma_{\rm adj}L^3}{|x|}=\frac{26.08{\,}{\rm GeV}^{-1}}{|x|}$,
to 2/3 (at $|x|\gg\frac{1}{\sqrt{\sigma_{\rm adj}}}=1.67{\,}{\rm GeV}^{-1}$). 
Clearly, $\frac{26.08{\,}{\rm GeV}^{-1}}{|x|}<\frac23$
up to $|x|=39.12{\,}{\rm GeV}^{-1}$, that is the distance much larger than $|x|_{\rm max}$. This proves our statement that
$\frac{1}{12^{1/3}}>\frac{L}{2}\left(\frac{\sigma_{\rm adj}}{|x|}\right)^{1/3}$.
As for the term $\frac{2}{3(\sigma_{\rm adj}x^2)^{2/3}}$, it exceeds $\frac{1}{12^{1/3}}$ only in the 
limiting case $\sigma_{\rm adj}x^2<1.89$, that is on the border of applicability of the inequality $\sigma_{\rm adj}x^2\gg 1$.
Thus, we conclude that, in general, the term $\frac{1}{12^{1/3}}$ is the defining one for the pre-exponential behavior of the 
function $D_1(x)$. This behavior is, therefore, $\sim x^{-2}$, that differs from the behavior $\sim |x|^{-1}$ of 
eq.~(\ref{D1S}).

\section{Green function of a two-gluon gluelump and the function $D$}
This gluelump is formed by two gluons, which are attached by fundamental strings to an infinitely heavy adjoint source and joined 
to each other by a third fundamental string. Since the gluons are identical, the vectors of their spatial coordinates, 
${\bf x}(\tau)$ and ${\bf y}(\tau)$, can be chosen in the form 
${\bf x}(\tau)=L'{\bf n}_1+{\bf l}_1(\tau)$, ${\bf y}(\tau)=L'{\bf n}_2+{\bf l}_2(\tau)$. Here, ${\bf n}_1=(\cos\theta, -\sin\theta, 0)$, 
${\bf n}_2=(\cos\theta, \sin\theta, 0)$, and $\theta$ is some constant angle, whose possible values as a function of $L'$ 
will be discussed at the end of the calculation.
The mean distance $L'$ from the gluons to the heavy source is, in general, different from $L$. 
Within such notations, the area swept out by the string 
joining the gluons is $2L'T\sin\theta$. After the extraction of the mean-area dependent factor, 
the scalar function $G^{(2{\rm gl})}(T,0)$, through which the 
Green function of the gluelump is expressed~\cite{S}, factorizes into the product of integrals over trajectories of each of the gluons:
$$
G^{(2{\rm gl})}(T,0)={\rm e}^{-2\sigma_{\rm fund}L'T\sin\theta}
\int\limits_{0}^{\infty}ds\int\limits_{0}^{\infty}ds'
\int\limits_{x_4(0)=0,{\,}x_4(s)=T}^{}{\cal D}x_4\int\limits_{y_4(0)=0,{\,}y_4(s)=T}^{}{\cal D}y_4
\int{\cal D}{\bf x}{\cal D}{\bf y}\times$$
\begin{equation}
\label{2g}
\times\exp\left[-\int_{0}^{s}d\tau\left(\frac{\dot{\bf x}^2}{4}+\frac{\dot x_4^2}{4}+\sigma_{\rm fund}|{\bf x}|\dot x_4\right)
-\int_{0}^{s'}d\tau\left(\frac{\dot{\bf y}^2}{4}+\frac{\dot y_4^2}{4}+\sigma_{\rm fund}|{\bf y}|\dot y_4\right)\right].
\end{equation}
Let us consider, for example, the integral over ${\bf x}(\tau)$. Substituting $|{\bf x}|\simeq L'+{\bf n}_1{\bf l}_1$, we have for it
$$
{\rm e}^{-\sigma_{\rm fund}L'T}
\int{\cal D}{\bf l}_1{\cal D}\mu
\exp\left[-\int_{0}^{s}d\tau\left(\frac{\dot{\bf l}_1^2}{4}+\frac{\mu}{2}+
\frac{\sigma_{\rm fund}^2\dot x_4^2({\bf n}_1{\bf l}_1)^2}{2\mu}\right)\right].$$
We can further use the same 
approximation as in the previous section, namely substitute $({\bf n}_1{\bf l}_1)^2\to\frac{1}{s}\int_{0}^{s}
d\tau({\bf n}_1{\bf l}_1)^2$. Together with the (free) integration over $l_1^3(\tau)$, this yields
$$
\frac{{\rm e}^{-\sigma_{\rm fund}L'T}}{\sqrt{4\pi s}}\int{\cal D}{\cal R}{\cal D}\mu\exp\left[-\int_{0}^{s}d\tau\left(
\frac{\dot{\cal R}^2}{4}+\frac{\mu}{2}+U({\cal R})\right)\right],$$
where ${\cal R}\equiv(l_1^1,l_1^2)$ and 
$$
U({\cal R})=
\left(\frac{\sigma_{\rm fund}^2}{2s}\int_{0}^{s}d\tau\frac{\dot x_4^2}{\mu}\right)(l_1^1\cos\theta-l_1^2\sin\theta)^2.$$
Using the variational method in the 2-d case (see~\cite{sb} for details), we further obtain 
$$
\int{\cal D}{\cal R}\exp\left[-\int_{0}^{s}d\tau\left(
\frac{\dot{\cal R}^2}{4}+U({\cal R})\right)\right]\simeq 
\frac{1}{4\pi s}\exp\left[-\frac{3}{\pi}\int d^2{\cal R}U({\cal R}){\rm e}^{-\frac{3{\cal R}^2}{s}}\right]=$$
$$=\frac{1}{4\pi s}\exp\left(-\frac{\sigma_{\rm fund}^2s}{12}\int_{0}^{s}d\tau\frac{\dot x_4^2}{\mu}\right).$$
The resulting integral over $x_4$ and $\mu$ then coincides (modulo the 
replacement $\sigma_{\rm adj}\to\sigma_{\rm fund}$) with that of 
eq.~(\ref{intl}), so that the total integral over $ds{\cal D}x_4{\cal D}{\bf x}$ in eq.~(\ref{2g}) reads
$$
\frac{(\sigma_{\rm fund}/T^4)^{1/3}}{3^{2/3}\cdot 2^{7/3}\cdot \pi^{3/2}}
\exp\left[-\sigma_{\rm fund}L'T-(9/32)^{1/3}(\sigma_{\rm fund}T^2)^{2/3}\right].$$
The integral over $ds'{\cal D}y_4{\cal D}{\bf y}$ apparently produces the same result. 
Taking into account the relation~\cite{S} 
$D(x)=\frac{g^4(N_c^2-1)}{2}G^{(2{\rm gl})}(x,0)$, we finally obtain that, at $\sigma_{\rm fund}x^2\gg 1$,
\begin{equation}
\label{Dfunc}
D(x)=\frac{g^4(N_c^2-1)}{3^{4/3}\cdot 2^{17/3}\cdot \pi^3}\left(\frac{\sigma_{\rm fund}}{|x|^4}\right)^{2/3}
\exp\left[-2\sigma_{\rm fund}L'(1+\sin\theta)|x|-
(9/4)^{1/3}(\sigma_{\rm fund}x^2)^{2/3}\right].
\end{equation}
Again, as in the case of the function $D_1(x)$, the term $2\sigma_{\rm fund}L'(1+\sin\theta)|x|$ in the exponent
is numerically larger than $(9/4)^{1/3}(\sigma_{\rm fund}x^2)^{2/3}$. Indeed, identifying  
$2\sigma_{\rm fund}L'(1+\sin\theta)$ with $M_0^{(2{\rm gl})}=2.56{\,}{\rm GeV}$ from the formula~\cite{S}
\begin{equation}
\label{DD}
D(x)\simeq\frac{g^4(N_c^2-1)}{2}0.108\sigma_{\rm fund}^2{\rm e}^{-M_0^{(2{\rm gl})}|x|}~ {\rm valid}~ {\rm at}~ M_0^{(2{\rm gl})}|x|\gg 1,
\end{equation}
we have $\frac{(9/4)^{1/3}(\sigma_{\rm fund}x^2)^{2/3}}{2\sigma_{\rm fund}L'(1+\sin\theta)|x|}\le
\frac{\left(9\sigma_{\rm fund}^2|x|_{\rm max}/4\right)^{1/3}}{M_0^{(2{\rm gl})}}=0.12$.
Furthermore, the mean distance between the two gluons, as well as the mean distance 
between these gluons and the heavy source may not exceed $|x|_{\rm max}$, i.e. the inequalities $2L'\sin\theta\le|x|_{\rm max}$, $L'\le|x|_{\rm max}$
should be respected together with the equation $2\sigma_{\rm fund}L'(1+\sin\theta)=M_0^{(2{\rm gl})}$. One can readily 
see that the unphysical case $\theta=0$
cannot be realized also numerically. Indeed, one should then have $\frac{M_0^{(2{\rm gl})}}{\sigma_{\rm fund}}=2L'\le2|x|_{\rm max}$, 
that is not feasible, since the l.h.s. equals $14.22{\,}{\rm GeV}^{-1}$, 
whereas the r.h.s. equals $11.12{\,}{\rm GeV}^{-1}$. For all other cases, we have: $\theta\le\arcsin\frac{|x|_{\rm max}}{2L'}$
at $\frac{|x|_{\rm max}}{2}\le L'\le|x|_{\rm max}$, while no restrictions on the values of $\theta$ 
exist at $L'<\frac{|x|_{\rm max}}{2}$.~\footnote{Note that $L'$ is bounded from below only by the inequality $L'>\frac{1}{M_0^{(2{\rm gl})}}$.}
Finally, we note that the pre-exponent in the function~(\ref{Dfunc}) falls off
as $|x|^{-8/3}$, whereas the pre-exponent of the function~(\ref{DD}) is a constant.

\section{Summary}
In the present paper, we have evaluated the functions $D(x)$ and $D_1(x)$, which parametrize the two-point gauge-invariant 
correlation function of gluonic field strengths in the stochastic vacuum model of QCD. To this end, we have applied the Feynman 
variational method to the calculation of the Green functions of one- and two-gluon gluelumps, through which the functions
$D$ and $D_1$ can be expressed. At the end of the calculation, we have compared our results with those found by the 
Hamiltonian analysis in ref.~\cite{S}. In case of the function $D_1$, this comparison yielded the mean separation of a gluon from the heavy source 
in the one-gluon gluelump. In case of the function $D$, it yielded some constraints on the values of the angle between the 
planar world sheets of the gluons in the two-gluon gluelump. We have further seen that, although fluctuations of the gluons 
(which are accounted for by the variational method) around their mean separations from the source produce the terms in the 
exponential asymptotics of the functions $D$ and $D_1$, which are parametrically larger than the leading ones, 
numerically these terms are nevertheless subleading, as they should be. Furthermore, we have also found the distance dependences 
of the pre-exponential factors in the both functions $D$ and $D_1$, which turn out to be different from those of ref.~\cite{S}. 
These differences are, however, of a rather academic interest, since numerically a distance dependence in the pre-exponent
plays an unimportant role at large distances considered both here and in ref.~\cite{S}.

In conclusion, the obtained results might be important for a field-theoretic establishment of the stochastic vacuum model. Besides that,
the performed calculation demonstrates how one can account for the fact that the virtual particles in QCD diagrams are confined rather than free.
Last but not least, the novel expressions for $D$ and $D_1$ obtained here might be useful for applications in high-energy scattering.

\section*{Acknowledgments}
This research was initially motivated by the discussions and collaboration with A.~Di~Giacomo, D.~Ebert,
E.~Meggiolaro, and Yu.A.~Simonov. Recently, I was encouraged by the discussions with N.~Brambilla and A.~Vairo, whom I also 
thank for informing me about some of the references. 
I would further like to thank the staff of the Physics Department of the Humboldt University of Berlin for the cordial hospitality.  
The work was supported by the Alexander~von~Humboldt foundation.

\end{document}